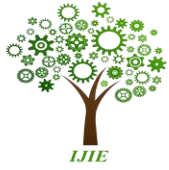

**International journal of innovation in Engineering**

journal homepage: www.ijie.ir

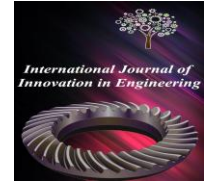

**Research Paper**

# A New Multi Objective Mathematical Model for Relief Distribution Location at Natural Disaster Response Phase

Mohammad Ebrahim Sadeghi[a], Morteza Khodabakhsh[b], Mahmood Reza Ganjipoor[b], Hamed Kazemipoor[b], Hamed Nozari[b]

[a] Department of Industrial Management, Faculty of Management, University of Tehran, Tehran, Iran

[b] Department of Industrial Engineering, Islamic Azad University, Central Tehran Branch

**ABSTRACT**

Every year, natural disasters such as earthquake, flood, hurricane and etc. impose immense financial and humane losses on governments owing to their unpredictable character and arise of emergency situations and consequently the reduction of the abilities due to serious damages to infrastructures, increases demand for logistic services and supplies. First, in this study the necessity of paying attention to locating procedures in emergency situations is pointed out and an outline for the studied case of disaster relief supply chain was discussed and the problem was validated at small scale. On the other hand, to solve this kind of problems involving three objective functions and complicated time calculation, meta-heuristic methods which yield almost optimum solutions in less time are applied. The EC method and NSGA II algorithm are among the evolutionary multi-objective optimization algorithms applied in this case. In this study the aforementioned algorithm is used for solving problems at large scale.

## 1- Introduction

Natural disasters cause, millions of casualties and leads to reduction of capabilities and financial losses across the world every year. As a developing country Iran is one of the geographical regions most prone to unpredictable emergency incidence and it is known as one of the ten most susceptible countries to natural disasters which almost 90 percent of its population are exposed to natural disasters. Considering the impact of natural disasters on the community's health, wellbeing, and welfare, providing healthcare services is the primary factor of survival and mortality reduction and individuals' welfare in the next phase after the occurrence of such incidents. Needless to say when encountering crisis and disasters, the initial action that will be taken is rescuing and mitigating the effects of the incident which requires quick response considering the limited time. Quick response which constitutes substantial part of crisis management, involves identification, assessment, decision making and provisional emergency measures, which whole part of this procedure takes place within a very short time sometime a few hours, therefor one of the actions that will be taken in order to manage the crisis is to consider relief measures aftermath of the crisis. The humanitarian environment of human rights organizations are built on three principles of humanity, impartiality and fairness (justice). In other words, they help any person who is in need of help, at any place regardless their beliefs even though in contradiction to theirs and do not privilege any group of final recipients over other beneficiaries (Webber, 1999). Humanity means that all the suffering and pain should be eliminated wherever it is. Fairness indicates help with no discrimination prioritized based on emergence of the needs. From physical point of view, the humanitarian environment indicates a peaceful area in which civilians and non-militant individuals and non-armed relief workers are able to freely move and operate. Often it is difficult to establish humanitarian principles in the complicated environments, especially in the situation of war and military conflict (Webber, 1999). Humanitarian logistics usually operates under a high level of uncertainty regarding demand and required supplies. Humanitarian workers often deal with many beneficiaries including aid workers, intermediaries, governments, and military beneficiaries and final recipient in a logistic humanitarian environment. The lack of coordination among humanitarian organizations in the crisis situation is common, as such there can be hundreds of humanitarian organizations with different political instructions, ideologies and religious beliefs aiming to carry out relief operations. The biggest challenge in these situations is the levelization of these organizations regardless of their beliefs and authorities (Farahani Hekmat-Far, 2009). The supply chain management concept is based on this concept that supply chains can outperform individual business units in the competition. A supply chain is consisting of a range of beneficiaries including suppliers, manufacturers, distributors, retailers and customers. The supply chain's structure varies for different companies even within similar industries. If the commercial supply chain is defined as a network streaming products, information and financial resources from the source to final customers, the relief or humanitarian supply chain therefore can be defined similarly as a network for managing the supplies, information and financial resources flow from relief sources to people affected by the crises. In the contrary to commercial supply chains which are owned by stockholders, the humanitarian supply chains are not centered around a targeted profit and they rely on help of volunteer forces and aid workers. Various models have been developed for establishing efficient supply chains based on cost minimization (which is equivalent of profit maximization) which can be used for establishing this kind of supply chains in their original forms or after some modification. An effective humanitarian supply chain management is capable of responding to many actions often on a global scale, as fast as possible and within a short timeframe. Thereby humanitarian supply chains must be multifold, universal, dynamic, and provisional in nature. Relief logistics has many actors, including a large number of private sectors, donors, and host governments and the military, governments of neighboring countries and NGOs, all of which have different political, ideological plans and religious attitudes, different levels of media attention and donors (Figure 1). When a crisis or a disaster occurs, depend on the location, type and intensity of the disaster different items and supplies are required. It is necessary to categorize the relief supplies due to their large quantity and variety which these categorizations will be very useful in identifying and evaluating required items and as well as designing the distribution system (Bozorgi Amiri et al., 2013). In one of these categorizations, relief items are segmented into two main categories of consumable and non-consumable (Figure 2). As demonstrated in Figure 2, the consumable relief items are distributed to the people affected by the disaster for several times, while non-expendable relief items require one-time distribution.

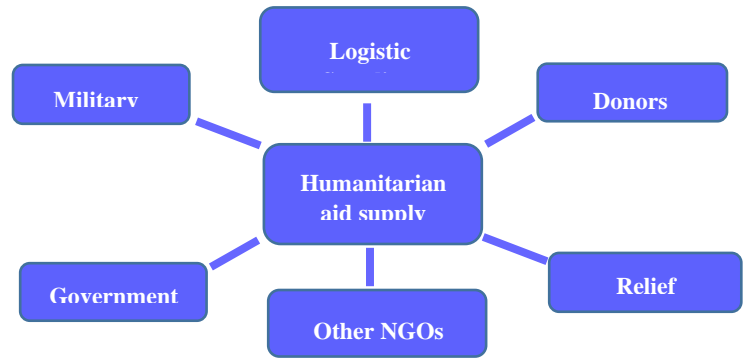

Figure 1: the actors involved in humanitarian supply chain (Rodriguez et al., 2004)

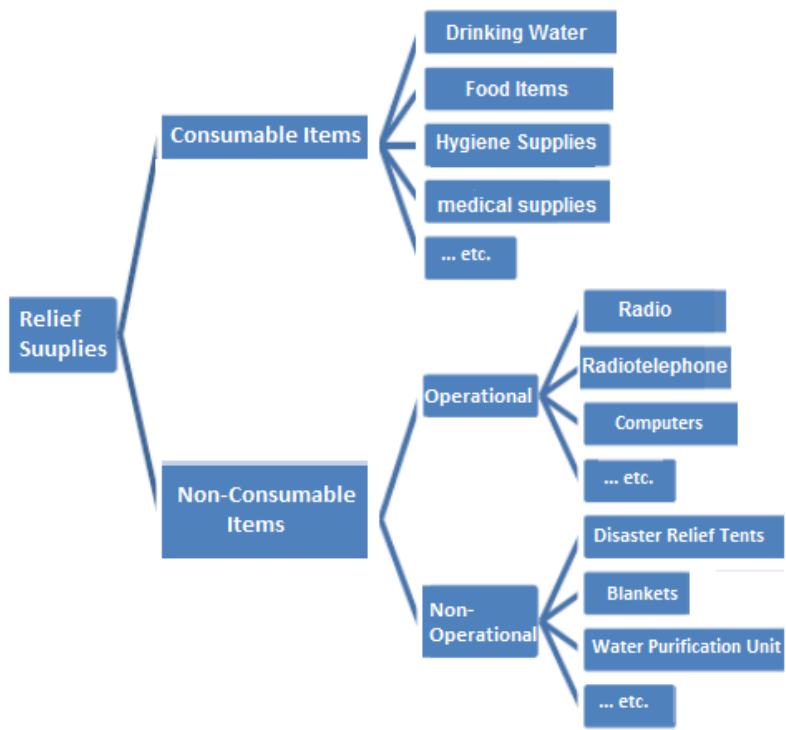

Figure 2: relief Item's segmentation (Murally et al., 2012)

Disaster Relief Logistics is one of the main activities in the subject of disaster management. The Relief Logistics means to distribute the right materials to people in the right places at the right time and to the right amount (Louveaux, 1993). The approaches applied include two steps: in the first step (preparedness phase), the location of relief distribution centers, the amount of inventory of relief supplies for storage purposes and for appropriate suppliers are determined. the next step (response phase) include decision making regarding the appropriate suppliers in the crisis occasions, the amount of relief supplies distributed from supply points to relief centers and from these centers to the areas affected by the disaster. Aid distribution and required services in a timely manner is one the important subjects in relief logistics (Bashiri et al., 2009). The summarized relief logistics operation is demonstrated in figure 3 (Bremberg & Joyale, 2001).

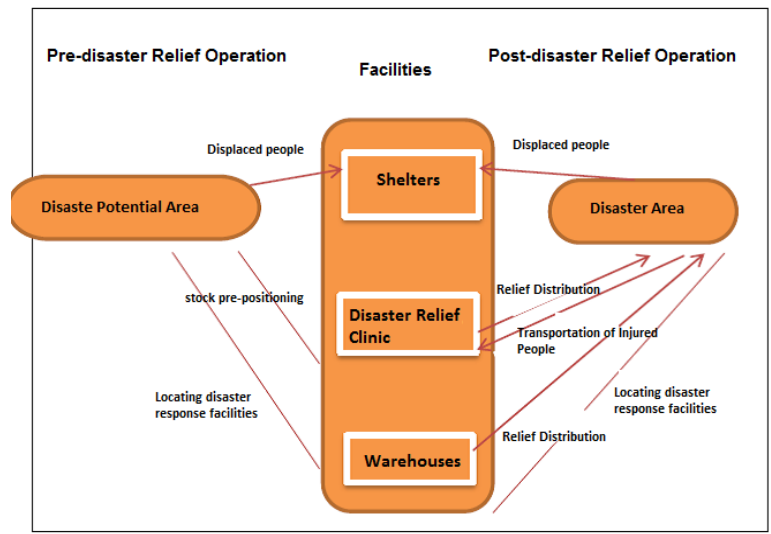

Figure 3: relief supply chain operation (Bremberg & Joyale, 2001)

## 2. Literature Review

Natural disasters and man-made disasters such as floods, earthquakes, hurricanes, droughts, wars and etc. affect people in different parts of the world every year. These natural crises (such as earthquakes, floods, hurricanes) owing to their unpredictable character cause immense and sometimes irreparable damage to nations. In fact, more than 599 disasters occur on our planet every year, kills nearly 75 thousand people and affect more than 200 million people (Bremberg & Joyale, 2001). For example, nearly 1,836 people lost their lives in the Katrina hurricane in 2005, and at least 86,000 people were killed in the Kashmir earthquake in the same year (Rodriguez et al., 2004; Levine and Arisil, 2004). Also, Haiti earthquake in 2012 affected nearly three million people. These natural disasters often cause huge financial and humane damages at time of occurrence. The top five disasters with the highest human casualties during a period from 1980 to 2012 is available in table 1 (Camacho Vallejo et al., 2014). Unfortunately, the natural and man-made disasters will increase to five times for the next 59 years, hence the need for relief services will rise as well. Thereby, it can turn the humanitarian aid delivery into a global industry. Most of the time data in disaster management are uncertain, and quick decisions are made based on just a fraction of the required information. Not only technology but also social, cultural and psychological factors play an important role in decision making (*kumawala*, 1972). Hence upon, this study proposes a Robust Possibilistic Programming, in crisis situations under uncertainty.

Table 1: The top five disasters with the highest human casualties during a period of 1980 to 2012 (Camacho Camacho Vallejo et al., 2014)

| Type | Number in the period | year | country | Number of mortalities | Number of affected people | Damages (million Dollars) |
|---|---|---|---|---|---|---|
| Earthquake | 756 | 2010 | Haiti | 222570 | 3700000 | 8000 |
|  |  | 2004 | Indonesia | 165708 | 532898 | 4451,6 |
|  |  | 2008 | China | 87476 | 45976569 | 85000 |
|  |  | 2005 | Pakistan | 73338 | 5128000 | 5200 |
|  |  | 1990 | Iran | 40000 | 710000 | 8000 |
| Hurricane | 2516 | 1991 | Bangladesh | 15000 | 1810000 | 50 |
|  |  | 2008 | Myanmar | 138366 | 2420000 | 2420000 |
|  |  | 1985 | Bangladesh | 15000 | 1810000 | 50 |
|  |  | 1998 | Honduras | 14600 | 2112000 | 3793,6 |
|  |  | 1998 | India | 9843 | 12628312 | 2500 |
| flood | 3120 | 1999 | Venezuela | 30000 | 483635 | 3160 |
|  |  | 1980 | China | 6200 | 67000 | 160 |
|  |  | 1998 | China | 3656 | 283973000 | 30000 |
|  |  | 1996 | China | 2775 | 154634000 | 12600 |
|  |  | 2004 | Haiti | 2665 | 31283 |  |

The disasters occurred during 1900 to 2012 with the highest casualties in Iran is shown in table 2. As it can be observed in the table, the type of disasters is mainly earthquake, due to presence of many active faults in Iran. Thus it can be concluded that Iran is an earthquake prone country.

Table 2: The worst disasters occurred in Iran according the number of casualties during 1900 to 2012 (Camacho Camacho Vllejo et al., 2014)

| Type of disaster | year | Number of casualties |
|---|---|---|
| Earthquake | 1990 | 40000 |
| Earthquake | 2003 | 26796 |
| Earthquake | 1978 | 25000 |
| Earthquake | 1962 | 12000 |
| Earthquake | 1968 | 10000 |
| Earthquake | 1972 | 5057 |
| Earthquake | 1909 | 5000 |
| Earthquake | 1929 | 3300 |
| Earthquake | 1957 | 3000 |
| Earthquake | 1930 | 2500 |

Altay (2006) proposed a model for crisis management centers location for relief supplies management. Jia et al. (2007) reviewed facility locating models for a large-scale emergency situation and categorized it into three groups: coverage models, median location and center location models. Yi and Kumar (2007) proposed a meta-heuristic algorithm (Ant colony optimization algorithm) to solve the disaster relief logistics problem. Balcik and Beamen (2008), proposed a model for determining the number and locations of distribution centers in rescue operations. They formulated the location problem of another type of maximum coverage problem for a set of probable scenarios. The proposed objective function maximizes the total expected demand covered by the distribution center. They also determined the amount of relief supplies required to be stored at each distribution center to meet the demand. Although they did not consider locating as part of the supply chain, their study is one of the first facilities locating problem in rescue operations that must be solved. Braldi and Bruni (2009) proposed a possiblistic model for determining the optimal location of emergency facilities under uncertainty. Table (3) demonstrates a summary of the proposed models by different researchers.

Table 3: Previous researches and their gaps and deficiency

| No. | researchers | year | Transportation vehicle capacity | | Distribution center capacity | | Aid delivery | | Solving method | |
|---|---|---|---|---|---|---|---|---|---|---|
| | | | With capacity | Without capacity | With capacity | Without capacity | Single layer | Multiple layers | exact | Meta heuristic |
| 1 | Sigal | 1999 | * | | * | | | * | * | * |
| 2 | Altay | 2006 | | * | | * | * | | * | |
| 3 | Tzeng et al. | 2007 | | * | * | | | * | * | * |
| 4 | Jia et al. | 2007 | | * | | * | | * | | * |
| 5 | Beamon & Balcik | 2008 | | | * | | | * | * | * |
| 6 | Braldi & Bruni | 2009 | * | | | * | | * | | * |
| 7 | Mete & Zabinsky | 2010 | | | * | | * | | * | |
| 8 | Lin et al. | 2011 | * | | * | | * | | * | |
| 9 | Boloori et al. | 2012 | * | | | | | * | | * |
| 10 | Haqhani & Afshar | 2012 | * | | * | | | * | * | * |
| 11 | Hirouki Ischi et al. | 2013 | | | * | | | * | | * |

| 12 | Farahani et al. | 2014 | | | | | | | | |
|----|-----------------|------|---|---|---|---|---|---|---|---|
| 13 | Camachou et al. | 2014 | * | | * | | | * | * | * |

## 3. Research Methodology

In this section, we propose a multi-objective mathematical model for the locating-distributing problem of relief items in the disaster response phase.

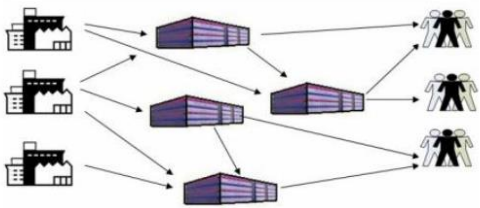

Figure 4: Mathematical planning model.

Model denotations are as follow:
- **I**  Set of the supply points (relief additive points)
- **J**  Set of the distribution centers for relief items
- **K**  Set of the affected points by the disaster
- **L**  Set of the relief supplies
- **S**  Sum of demand quantity (indicates the amount of demand for each relief supplies, e.g. the amount of food, medicine or water and etc.)
- **D**  Sum of demand numbers (indicates the number of demands, in other word the number of people affected by disaster who demand relief supplies).
- **M**  Transportation vehicles
- **T**  Type of the equipment
- **X**  Traveled distance

Model parameters
- $F_j$   The cost of setting up the relief distribution center j
- $d_{mk}$  The amount of relief needed from the m vehicle at the damaged point k
- $S_{mi}$  The amount of goods collected from the vehicle m at the supply point i
- $C_{ij}$  The cost of transferring each unit from the supply point i to the relief distribution center j
- $C_{jk}$  The cost of transferring each item from the relief distribution center j to the damaged point k
- $\pi_k$  The cost of defective goods per damaged point k

- $N_{ij}$    Total capacity used by the vehicle from the supply point i to the relief distribution center j
- $N_{jk}$    The total capacity used by the vehicle from the relief distribution center j to the damaged point k
- $f_{ks}$    The cost of facilitating the s level at the candidate's place k
- $P_N$    Number of Candidate Points
- $r_{dj}$    Distance between j base and demand point d
- $r_{djt}$    The distance between j base and demand point d for each t type equipment
- $R_t$    T coil coverage radius
- $V_t$    Average speed of t type equipment

Model variables

- $x_{jk}$    The distance traveled from the distribution center j to the new distribution center k of the candidate k (represents the distance traveled from the distribution center to the new distribution point serving the affected area in the new area)
- $x_{dj}$    The distance traveled from the demand point d to the distribution center j
- $x_{dk}$    The distance traveled from the demand point d to the damaged point k
- $x_{jj'}$    The distance traveled from the distribution center j to the higher level distribution center j
- $Q_{mij}$    The amount of goods transferred by the vehicle m from the supply point i to the relief distribution center j
- $Y_{mjk}$    The amount of goods transferred by the vehicle m from the distribution center j to the affected area k
- $Y_{ks}$    If, 1 in the candidate's place, k is facilitated by the surface s and zero is otherwise.
- $Y_{jklm}$    Equivalent to the amount of cargo l of the product, using the transport m from the storage center j to the k region
- $Z_j$    If the relay distribution center j is selected to be opened, it is equal to 1 and otherwise 0
- $T^1_{ijlm}$    When a product l is required to receive a shipment of m from a donor country or an international organization i to a storage facility j.
- $T^2_{jklm}$    The time needed to distribute a ll product by using m transport from the storage center j to the damaged area k
- $D_{djs}$    The share of demand for the surface s of the demand point d, which is met by the distribution center j
- $D_{dks}$    A share of the demand level s of the demand point d that is satisfied by the affected center k.
- $D_{jks}$    A share of s level demand, which is referred from the distribution center j to k (ie, the level s of the demanded amount of the injured)
- $Y_{ks}$    If, at the candidate location, k is the center of the distribution of the surface s, then zero is set to zero otherwise
- $F_j$    It is equal to one when the distribution center is established at j, otherwise zero Gets

- $U_{dkt}$ Is equal to one that is assigned the type t at the point of demand d to the damaged point k , Otherwise it becomes zero.

The objective function is described as follow:

$$\min f_1 = \sum_{m,k} \pi_k \left( d_{m,k} - \sum_j y_{mjk} \right) \tag{1}$$

$$\min f_2 = \sum_i \sum_j \sum_s c_{ij} x_{ij} D_{djs} + \sum_i \sum_k \sum_s c_{ik} x_{ik} D_{dks} + \sum_j \sum_k \sum_{t \in S | t \geq s} jk \, x_j D_{jks} + \sum_k \sum_s f_{ks} y_{ks} \tag{2}$$

$$\min_{y_{jklm} x_{ijlm}} f_3 = \sum_{i \in I} \sum_{j \in J} \sum_{l \in L} \sum_{m \in M} T^1_{ijlm} + \sum_{j \in J} \sum_{k \in K} \sum_{l \in L} \sum_{m \in M} T^2_{jklm} \tag{3}$$

**Objective 1** minimizes the total cost of supplies shortage.

**Objective 2** is the total cost of delivering demanded supplies and the cost of constructing the distribution center, which includes four group of costs as follow: the first group of delivering demanded supplies between customers and the existing distribution center, second group is delivering demanded supplies between customers and the recently established distribution center, the third grope is delivering demanded supplies between distribution centers and the areas affected by the disaster. The fourth group consists of fixed expenses.

**Objective 3** minimizes the total response time which is defined as the required time to deliver the aid.

**Constraints:**

$$s.t. \sum_{k \in K} \sum_{l \in L} \sum_{m \in M} v_l y_{jklm} \leq V_t \quad \forall m \tag{4}$$

$$\sum_{j,k} y_{mjk} \leq \min \left\{ \sum_k d_{m,k}, \sum_i s_{m,j} \right\} \cong \quad \forall m \tag{5}$$

$$r_{dj} r_{djt} \leq R_t \quad \forall d,j,t \tag{6}$$

$$T^1_{ijlm} = \begin{cases} t^1_{ijlm} \; if \, x_{ijlm} \geq 1 \\ 0 \end{cases} \quad \forall m \tag{7}$$

$$T^2_{jklm} = \begin{cases} t^2_{jklm} \; if \, x_{jklm} \geq 1 \\ 0 \end{cases} \quad j \in J \tag{8}$$

$$\sum_{j,k} y_{mjk} \leq \lambda \quad \forall m \tag{9}$$

$$\lambda \leq \sum_k d_{m,k} \quad \forall m \tag{10}$$

$$\lambda \leq \sum_i s_{m,i} \quad \forall m,j \tag{11}$$

$$\sum_{i} Q_{mij} = \sum_{k} y_{mjk} \qquad \forall m,i \qquad (12)$$

$$\sum_{j} Q_{mij} \leq s_{mi} \qquad \forall m,k \qquad (13)$$

$$\sum_{j} y_{mjk} = d_{mk} \qquad \forall m,i,j \qquad (14)$$

$$Q_{mij} \leq N_{ij} z_j \qquad \forall m,j,k \qquad (15)$$

$$Y_{mjk} \leq N_{jk} z_j \qquad \forall m,i,j \qquad (16)$$

$$Q_{mij} \geq 0 \qquad \forall m,j,k \qquad (17)$$

$$y_{mjk} \geq 0 \qquad \forall j \qquad (18)$$

$$z_j \in \{0,1\} \qquad \forall s \qquad (19)$$

$$\sum_{k} y_{ks} \leq P_N \qquad (20)$$

### 3.1. The procedure of presenting a model for demand uncertainty

According the presented model in this section, different fuzzy approaches to planning are considered.

**BPCCP model**

For modeling the objective function, the operator's expected value and the appropriate value that will fulfill the possible constraints with vague parameters were used. Also, we considered the trapezoidal probability distribution with four outstanding points as $\xi = \big(\xi(1), \xi(2), \xi(3), \xi(4)\big)$ for the vague parameters (Fig. 5).

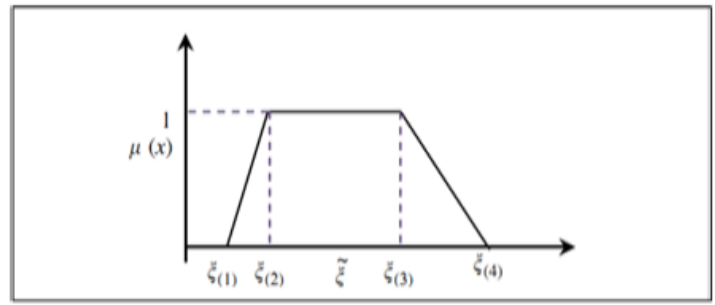

Figure 5: A trapezoidal fuzzy number.

**RPP model**

First, we assume that the f, c, and d vectors are ambiguous parameters, and in this phase the coefficient matrix N are not vague. Therefore, the BPCCP model is defined as the RPP model. This model aim to establish balance between the sections of the objective function:

1. average performance
2. optimality robustness

3. Feasibility robustness

In this model same as the BPCCP model, the first term in the objective function indicates the expected value for z and the second term indicates the difference between two final possible value for z which are defined as follow:

$$Z_{max} = f_4 x + c_4 x \tag{21}$$

$$Z_{min} = f_1 x + c_1 x \tag{22}$$

And parameter $\partial$ indicates importance (weight) of this term against other two terms of the objective function. Thus, the second term ensures that the gap between the maximum and minimum expected optimal value Z of the objective function will be obtained. In fact, this term controls the optimality robustness of the solution vector. The third term determines the confidence level of any possible constraint, and the penalty for any possible violation of any constraint, including ambiguous parameters, and shows the difference between the worst possible value of the uncertain parameter and its value in the chance constraints; in fact, it controls the feasibility robustness in the solution vector. It should also be noted that it is not only a theoretical or meaningless parameter, but it can be determined based on application in the right context.

Although here the BPCCP model is the minimum of the confidence level of possible constraint variable and its value is optimized based on the objective function, thus the RPP model avoid subjective judgments about the best value of the level of confidence level of possible constraints to understand the global optimum value of them, and also manages the uncertainty of the parameters, and as a result it can be classified as a method. In addition, the model can easily yield the optimum value for confidence levels of increasing number of possible constraints needless to sophisticated application and time-consuming processes, such as simulation experiments based on disaster description.

$$5.0 \leq \wp, \Im, , \eta, \vartheta, \psi > 1 \tag{23}$$

Now let's assume that the coefficients matrix N is uncertain, and M is a very large number, and same as the previous model RPP-I it is defined as follow:

$$\min \left(\frac{(F_1+F_2+F_3+F_4)}{4}\right) Q_{mij} + \left(\frac{(c_1+c_2+c_3+c_4)}{4}\right) y_{mjk} + \left(\frac{a_1+a_2+a_3+a_4}{4}\right) z_j + \partial\{(F_4 Q_{mij} + c_4 y_{mjk} + a_4 z_j) - (F_1 Q_{mij} + c_1 y_{mjk} + a_1 z_j)\} + \wp\left((1-\alpha)s_{2mi} + \alpha s_{1mi} - s_{1mi}\right) + \Im\left((1-\beta)d_{2mk} + \beta d_{1mk} - d_{1mk}\right) + \hbar\left((1-\mu)N_{2ij} + \mu N_{1ij} - N_{1ij}\right) z_j + \eta\left((1-\rho)N_{2jk} + \rho N_{1jk} - N_{1jk}\right) + \vartheta\left((1-\phi)d_{2mk} + \phi d_{1mk} - d_{1mk}\right) + \psi\left((1-\varphi)s_{2mk} + \varphi s_{1mk} - s_{1mk}\right) \tag{24}$$

$$\max \tilde{f}_2 = \sum_w w_m \tag{25}$$

$$\frac{\sum_j y_{mjk}}{w_m} \geq k d_{4mk} + (1-k) d_{3mk} \tag{26}$$

$$\min \tilde{f}_1 = \left(\frac{d_1 + d_2 + d_3 + d_4}{4}\right) \pi_k - \sum_{m,k} \pi_k \left(\sum_j y_{mjk}\right) \tag{27}$$

$st: x, y, z, \alpha, \beta, \rho, \varphi \in B$

As you can see in the model, when technical coefficients are assumed uncertain, the linearity of the constraints and the possible objective function will turn to nonlinear, although in some cases a nonlinear term can be turned into linear with defining a new variable and adding few constraints to the model. To avoid the complexity of the nonlinear model, the auxiliary variable and the nonlinear model will be turned into an equivalent linear model.

$$\mu z_i = v \tag{28}$$

$$\rho z_j = \varsigma \tag{29}$$

Thereby, the above nonlinear model will turn into the following linear model:

$$min\left(\frac{(F_1+F_2+F_3+F_4)}{4}\right)Q_{mij} + \left(\frac{(c_1+c_2+c_3+c_4)}{4}\right)y_{mjk} + \left(\frac{a_1+a_2+a_3+a_4}{4}\right)z_j + \partial\{(F_4Q_{mij} + c_4y_{mjk} + a_4z_j) - (F_1Q_{mij} + c_1y_{mjk} + a_1z_j)\} + \wp((1-\alpha)s_{2mi} + \alpha s_{1mi} - s_{1mi}) + \Im((1-\beta)d_{2mk} + \beta d_{1mk} - d_{1mk}) + \hbar\left((z_j - v)N_{2ij} + vN_{1ij} - z_jN_{1ij}\right)z_j + \eta\left((z_j - \varsigma)N_{2jk} + \varsigma N_{1jk} - z_jN_{1jk}\right) + \vartheta((1-\phi)d_{2mk} + \phi d_{1mk} - d_{1mk}) + \psi((1-\varphi)s_{2mk} + \varphi s_{1mk} - s_{1mk}) \tag{30}$$

## 4. Results and Discussion

In this section, a solution for the problem is discussed and furthermore, analysis and interpretation of the results are presented. The NSGA II meta-heuristic algorithm was used to solve the problems, and the computer programming was conducted with MATLAB R2010a and was carried out using a portable computer with CPU: 2.30 GHz and 4 GB of RAM. Also, to generate different values for the two objective functions that have been defined in this problem a fuzzy numerical simulation algorithm was applied, which is described as follow. Furthermore, a neural network system with a fuzzy inference engine is designed for the problem.

### 4.1. Problem Solving Method

Given the multi objective nature of this mathematical model, a NSGAII Meta heuristic algorithm is used and the results are presented. Also, to achieve an efficient objective function, we designed a fuzzy neural network to find the Pareto solution, which will be fully discussed in this section. It is worth noting that a fuzzy simulation algorithm was used to generate different values for the objective function problems.

### 4.2. The NSGAII Algorithm

The NSGAII algorithm which uses non-dominant sorting is one of the most well-known and widely used optimizing algorithms for multi-objective optimizing. This algorithm was first proposed by Deb in 2002. Alongside all the application that is provided by the NSGAII has, it can be considered as the formation pattern of many multi-objective optimization algorithms. This algorithm and its unique approach to multi-objective optimization problems have been frequently used to develop new multi-objective optimization algorithms.

### 4.3. Comparative Metric for Evaluating Multi-objective Algorithms

There are two main metric groups to evaluate the performance of multi-objective meta-heuristic algorithms, including convergence metrics and dispersion metrics.

The first metrics group is number of Pareto solution and Mean of Ideal Distance, the second group metrics are spacing and Diversification and also the time metrics. According to work of Bashiri, 2009 and Afshar, 2011, four metrics was applied to compare these algorithms in this study, which are explained in the future.

**The Mean of Ideal Distance Metric**

This metric is used for calculating the average distance from Pareto solutions to the origin of coordinates. According to this equation, it is obvious that the lower this metric is, the greater the efficiency of the algorithm will be. Since, one of the goals of multi-objective Pareto-based approach is to have far closer solution to the origin boundaries, thus this metric calculates the distance of the fronts from the best value for the population. The n is the numbers of Pareto solutions and the mean distance from ideal points is obtained as the value for this metric. Because the type of objective function is minimization with equal weight for each objective, we selected the origin point as a best value.

In this thesis, considering the fact that the both objective functions are minimization functions, thus the ideal point is equal to the minimum of each of the objective functions in all the algorithms. The MID index equation is used for calculation.

$$MID = \sum_{i=1}^{n} \frac{\sqrt{\left(\frac{f_{1i} - f_{1.i}^{best}}{f_{1.total}^{max} - f_{1.total}^{min}}\right)^2 + \left(\frac{f_{2i} - f_{2.i}^{best}}{f_{2.total}^{max} - f_{2.total}^{min}}\right)^2}}{n} \quad (31)$$

**Algorithm running time metric (CPU T)**

In the large scale problems one of the important indexes is the solution running time, therefore algorithm running time is considered as one of the quality assessment indexes.

**Spacing metric (SM)**

This metric represent the uniformity of Pareto solution distribution in the solution settings. This metric is calculated as follow: The lower the SM index, the higher performance of the algorithm.

$$SM = \frac{\sum_{i=1}^{n-1}|\bar{d} - d_i|}{(n-1)\bar{d}} \quad (32)$$

**Diversification metric (DM)**

The diversification metric measures the diversity range of the Pareto solutions of an algorithm and it is calculated according the following equation. The higher the DM index, the higher performance of the algorithm.

This parameter shows the width of the Pareto responses of an algorithm and can be calculated by the following equation. The higher the DM index, the better the algorithm.

$$DM = \sqrt{\left(\frac{\max f_{1i} - \min f_{1i}}{f_{1.total}^{max} - f_{1.total}^{min}}\right)^2 + \left(\frac{\max f_{2i} - \min f_{2i}}{f_{2.total}^{max} - f_{2.total}^{min}}\right)^2} \tag{33}$$

**Experiment design to set algorithm parameters**

As it was discussed in the previous sections, the NSGAII algorithm involves many factors and parameters, affecting the final results and performance of the algorithm, therefore the algorithm performance is vastly affected by the appropriate combination of these factors. One of the objectives of experimental design is to observe and identify the shift in the output variables corresponding to intended shifts made to the input variables.

There are several methods to design an experiment. One of the first methods presented in this field is the factorial method which obtains the number of experiments by the equation $= L^m$. The main disadvantage of this method is that if there are many variables, it results in large number of experiments and this is not economically efficient in terms of time and costs.

Therefore, it is necessary to find a way to reduce the numbers of the experiments. One of these modifications is the Taguchi method.

This method can be summarized in four general steps:

1. Presenting the factors effective on reaction
2. Required number of the experiments
3. Solutions analysis
4. Assessment of the optimum condition

In the first step, the effective factors are specified and several modes are assigned to each one. The number of experiments is determined by the number of effective parameters and the number of levels they have (different values assigned to each parameter during the experiment). After determining the number of experiments, a matrix is defined, which the rows of this matrix indicate the conditions of the experiment. In this approach, it is expected that the solutions analysis yields the following results:

- Optimal conditions which yield optimal quality
- The extent of which the factors affect the performance and quality.

There are two methods for analyzing the experiment:

1- Standardized method of ANOVA (Analysis of variance)
2- use of signal-to-noise ratio (S/N)

The Taguchi method classifies the objective functions into three categories: the lower the better, the higher the better, and the best nominal. Since our objective function is intended to minimize the prediction error, it is considered to be in the first categories. In this research, the proper values for the algorithm parameters are determined during the sequential steps with the help of Minitab software.

Table 4 demonstrates the ranges for important parameters of the NSGAII algorithm.

Table 4: The ranges of the NSGAII algorithm parameters

| Factors | Levels |
|---|---|
| **Crossover rate** | 0.9, 0.75, 0.65 |
| **Mutation rate** | 0.30, 0.20, 0.10 |
| **Initial population** | 150, 100, 50 |

Now, in order to determine the optimal parameters, the criteria are converted to a solution using one of the multiple attribute decision making methods, called Simple Additive Weighting (SAW). Accordingly, the instance problems were generated and solved for all of the 22 levels, and the results were imported to the Minitab software for the Taguchi experimental design. The settings of parameters resulted from solving the instance problems and implementation of the aforementioned steps are depicted in the figure 6.

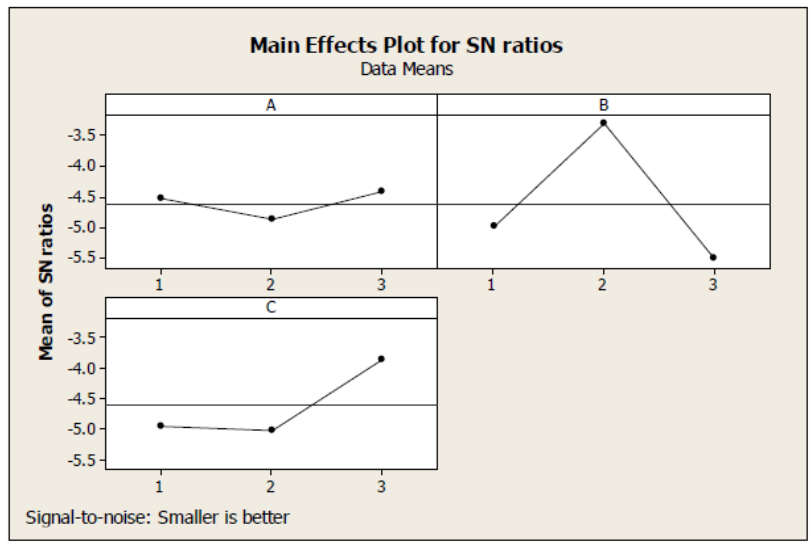

Figure 6: Parameter settings of NSGA-II Algorithm

The optimum combination for the parameters of the NSGAII algorithm can be set by examining Figure 7. Table 5 shows these values. According to table 5, values including 0.9 for crossover probability, 0.2 for mutation and 150 for initial population were set. Also, according to the initial experimental data, the number of generations required to improve the algorithm is set as 50.

Table 5: Proposed parameters value for the NSGAII algorithm

| Number of generation | Initial population | Mutation rate | Crossover rate |
|---|---|---|---|
| 50 | 150 | 0.2 | 0.1 |

### 4.4. Problem Solving using the GAMS software

The operational research software, GAMS is a high-level programming language used to solve mathematical programming models. This software is a very high-speed problem solver used for large scale models. In fact, GAMS can be considered as the best software for solving the most sophisticated and large scale optimization problems. The GAMS software easily connects to databases. Also, if problem doesn't involve high volume of input data, then the data can be entered into the software directly. For modeling, all variables and constraints are set once. Since the software easily iterate restrictions and there is no need to insert all the restrictions. The GAMZ software output is easy to use. The outputs can be specified in the software itself, or it can be for example, imported to the Excel after connecting to the databases. In this thesis 22 problem instances are solved using GAMS software, which the results of solving the problems are shown in the table (6).

Table 6: Results of solving problem instances in GAMS software

| Problem No. | $i$ | $j$ | $k$ | $s$ | $Z_1$ | $Z_2$ | $Z_3$ |
|---|---|---|---|---|---|---|---|
| 1 | 3 | 1 | 2 | 3 | 6031 | 3254348 | 5 |
| 2 | 3 | 2 | 3 | 3 | 8522 | 34655784 | 7 |
| 3 | 4 | 2 | 2 | 2 | 5409 | 2876452 | 3 |
| 4 | 4 | 2 | 3 | 3 | 9477 | 3542456 | 7 |
| 5 | 5 | 2 | 3 | 2 | 8028 | 2356481 | 5 |
| 6 | 5 | 3 | 3 | 3 | 9208 | 5265498 | 8 |
| 7 | 5 | 3 | 4 | 4 | 14896 | 3546751 | 10 |
| 8 | 8 | 4 | 5 | 3 | 15434 | 4568795 | 13 |
| 9 | 8 | 4 | 6 | 4 | 19670 | 5643579 | 21 |
| 10 | 10 | 4 | 5 | 3 | 17847 | 4325468 | 13 |
| 11 | 10 | 5 | 6 | 3 | 21930 | 4167954 | 15 |
| 12 | 15 | 7 | 9 | 3 | 32506 | 3648792 | 24 |
| 13 | 15 | 8 | 11 | 3 | 37112 | 6698627 | 29 |
| 14 | 15 | 9 | 12 | 4 | 44402 | 4568938 | 41 |
| 15 | 25 | 12 | 15 | 4 | 71023 | 5426579 | 52 |
| 16 | 50 | 25 | 20 | 4 | 151245 | 3698657 | 88 |
| 17 | 80 | 30 | 40 | 4 | 239208 | 3657948 | 145 |
| 18 | 100 | 40 | 55 | 4 | 386627 | 4542546 | 145 |
| 19 | 150 | 60 | 90 | 4 | 759371 | 6336597 | 330 |
| 20 | 250 | 100 | 140 | 4 | 1210974 | 823649 | 518 |
| 21 | 400 | 150 | 250 | 4 | 3202195 | 6556249 | 932 |
| 22 | 500 | 200 | 300 | 4 | 5343533 | 4754682 | 1119 |

### 4.5. Random instance generation

In order to verify the proposed model and the efficiency of the proposed algorithm, five problems are designed in different dimensions, which their related information are demonstrated in table (7). Also the specified bounds for the problem parameters are demonstrated in table (8).

Table 7: the number of different levels of the problem instances

| problem | I | J | K | L | S | D | M | T |
|---|---|---|---|---|---|---|---|---|
| 1 | 1 | 1 | 2 | 3 | 3 | 5 | 2 | 2 |
| 2 | 2 | 3 | 2 | 3 | 5 | 2 | 2 | 2 |
| 3 | 3 | 4 | 5 | 5 | 2 | 10 | 3 | 3 |
| 4 | 5 | 2 | 1 | 1 | 15 | 12 | 5 | 5 |
| 5 | 2 | 12 | 12 | 14 | 20 | 25 | 2 | 2 |

In table (7), the first column shows the problem numeral, the second column shows the number of supply points (first level), the third column shows the number of distribution centers, the fourth column shows the affected areas (second level), the fifth column represents the number of relief supplies, the sixth column shows the amount of demand, and the seventh column shows the quantity of demand, column eighth shows the number of transportation vehicles, an Eighth d the ninth column represents the type of equipment.

Table 8: input parameters distribution for the problem instances.

| Values | Parameters | Values | Parameters |
|---|---|---|---|
| Uniform (5.7) | $F_j$ | Uniform (1, 2) | $x_{dk}$ |
| Uniform (5.7) | $d_{mk}$ | Uniform (2, 3) | $x_{jj}$ |
| Uniform (5.7) | $S_{mi}$ | Uniform (4, 8) | $e_{ks}$ |
| Uniform (5.7) | $C_{ij}$ | Uniform (20, 26) | $P_N$ |
| Uniform (5.7) | $C_{jk}$ | Uniform (20, 26) | $r_{dj}$ |
| Uniform (5.7) | $\pi_k$ | Uniform (20, 26) | $r_{djt}$ |
| Uniform (50, 80) | $N_{ij}$ | Uniform (20, 26) | $r_{dh}$ |
| Uniform (50, 80) | $N_{jk}$ | Uniform (20, 26) | $R_t$ |
| Uniform (50, 80) | $x_{jk}$ | Uniform (20, 26) | $V_t$ |
| Uniform (50, 80) | $x_{dj}$ | Uniform (20, 26) | $T_d$ |

Also in the table (8), the required parameters of this problem were randomly generated using uniform distribution at proper time intervals. After the random generation of the instances in various dimensions, the problems were implemented in the GAMS software, and these problems were ran in different sizes with proposed values, and the result are demonstrated in the following tables.

### 4.6. Examining the proposed NSGAII algorithm efficiency and solving the problem instances

In the proposed ε-constraint Method, the first objective function is considered as the primary objective function and the second one as the secondary objective function; next, 10 breakpoints were specified for the second objective function, and a total of 10 Pareto points for each problem is generated. In this section, the five-dimensional problems designed in the previous sections, notwithstanding both objective functions, are solved using the ε-constraint Method and the proposed NSGA II algorithm, and the resulted Pareto solutions are reported. In order to explain the ε-constraint Method, we consider the No. 3 problem. First of all, in order to set the Pareto points based on each objective function the problems are solved separately. The results are reported in Table (9).

| Problem No. | Type of objective | f1 | f2 | f3 |
|---|---|---|---|---|
| 3 | Min f1 | 164 | 0 | 1015 |
|   | Min f2 | 178 | 1 | 929 |
|   | Min f3 | 923 | 0 | 854 |

After specifying the values of table (9), according to the third step of the ε-constraint Method, 10 ε values are assigned to the second objective function. Next, in table (10) the ε values for different breakpoints are calculated for the second to third objective functions. The 10 breakpoints are assigned.

Table 10: The resulted different ε values and the objective function values for problem No. 3

| No. | Epsilon value ($\varepsilon_2$) | Epsilon value ($\varepsilon_3$) | $Obj_1$ | $Obj_2$ | $Obj_3$ |
|---|---|---|---|---|---|
| 1 | 0.1 | 861.9173 | 307 | 0.34 | 1302 |
| 2 | 0.2 | 869.4276 | 358 | 0.40 | 1280 |
| 3 | 0.3 | 876.9379 | 365 | 0.48 | 1229 |
| 4 | 0.4 | 884.4482 | 388 | 0.53 | 1140 |
| 5 | 0.5 | 891.9585 | 398 | 0.64 | 992 |
| 6 | 0.6 | 899.4688 | 437 | 0.66 | 957 |
| 7 | 0.7 | 906.9791 | 444 | 0.78 | 930 |
| 8 | 0.8 | 914.4894 | 456 | 0.80 | 876 |
| 9 | 0.9 | 921.9997 | 458 | 0.89 | 870 |

Next, the Pareto frontiers resulted using the two algorithm for the problem No.3 is shown in table 11. It is evident that NSGA II algorithm has successfully resulted 9 Pareto solutions for the problemNo.3.

Table 11: Pareto-optimal solutions resulted by solving the problem instance No.3 using EC and NSGA II.

| No. | EC | | | NSGA II | | |
|---|---|---|---|---|---|---|
|  | First Objective | Second Objective | Third Objective | First Objective | Second Objective | Third Objective |
| 1 | 1302 | 0.34 | 307 | 1408 | 0.26 | 324 |
| 2 | 1280 | 0.40 | 358 | 1391 | 0.29 | 354 |
| 3 | 1229 | 0.48 | 365 | 1352 | 0.36 | 367 |
| 4 | 1140 | 0.53 | 388 | 1219 | 0.43 | 371 |
| 5 | 992 | 0.64 | 398 | 1211 | 0.56 | 393 |
| 6 | 957 | 0.66 | 437 | 1079 | 0.59 | 404 |
| 7 | 930 | 0.78 | 444 | 1028 | 0.66 | 418 |
| 8 | 876 | 0.80 | 456 | 982 | 0.85 | 427 |
| 9 | 870 | 0.89 | 458 | 912 | 0.89 | 430 |
| 10 | 861 | 0.94 | 466 | - | - | - |

In Figures (7) to (9), the resulted Pareto points using the two algorithms for problem No.3 are demonstrated as an example in small scale. It is noteworthy that these points and objective functions are shown correspondingly.

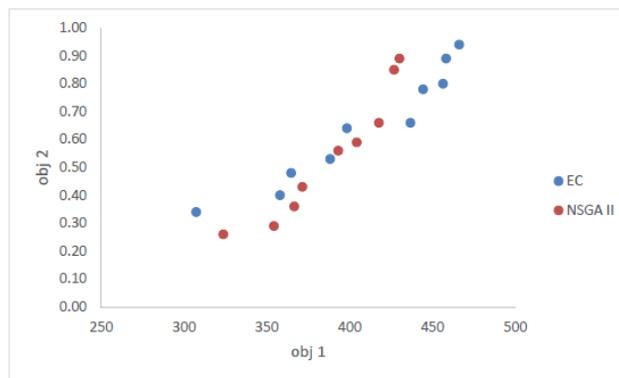

Figure 7: The resulted Pareto frontiers in Problem instance No. 3 for the first and second objective functions

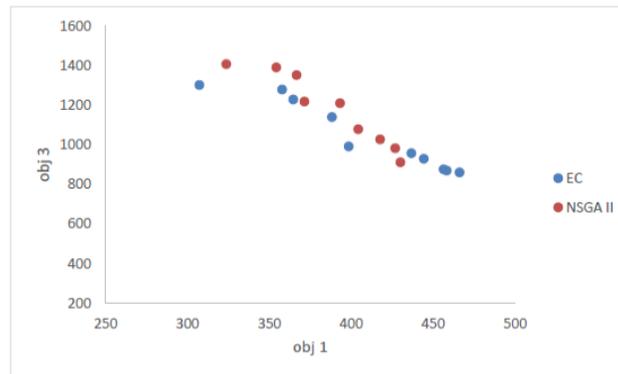

Figure 8: The resulted Pareto frontiers in Problem instance No. 3 for the first and third objective functions

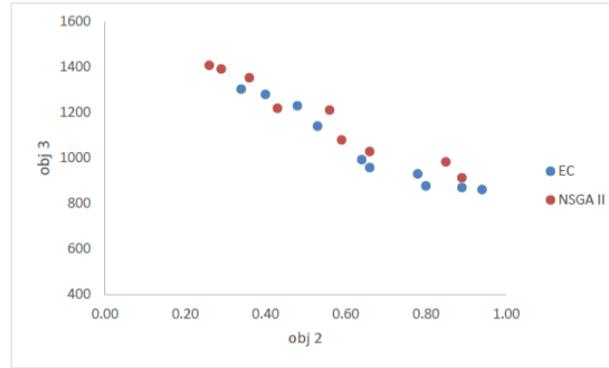

Figure 9: The resulted Pareto frontiers in Problem instance No. 3 for the second and third objective functions

According to Figures (8) to (10), it is evident that the proposed Pareto frontiers resulted from the NSGA II algorithm is very close to the frontiers resulted using the exact ε-constraint Method. Although in order to verify the proposed algorithm more exact, and to what extent this algorithm is able to identify the Pareto optimal frontier, the metrics presented in the previous sections are applied. For this purpose, the three MID, SM, and DM metrics are calculated, and according to SAW values resulted from this three metrics, the proposed algorithm performance is examined. The calculated values for the resulted frontiers using the two proposed algorithms from the problem instance No.1 to No.5 are reported in the tables (12) to table (16).

Table 12: The NSGA II Algorithm verification for the problem No.1

| Method/metric | MID | SM | DM | SAW |
|---|---|---|---|---|
| ε-constraint | 0.93 | 0.95 | 1.56 | 1.23 |
| NSGA II | 0.97 | 0.98 | 1.22 | 1.09 |

Table 13: The NSGA II Algorithm verification for the problem No.2

| Method/metric | MID | SM | DM | SAW |
|---|---|---|---|---|
| ε-constraint | 0.78 | 0.66 | 1.62 | 1.47 |
| NSGA II | 0.81 | 0.73 | 1.47 | 1.36 |

Table 14: The NSGA II Algorithm verification for the problem No.3

| Method/metric | MID | SM | DM | SAW |
|---|---|---|---|---|
| ε-constraint | 1.14 | 0.74 | 2.41 | 1.55 |
| NSGA II | 1.15 | 0.82 | 2.21 | 1.43 |

Table 15: Output Results of NSGA II Algorithm for the problem No.4

| Method/metric | MID | SM | DM | SAW |
|---|---|---|---|---|
| ε-constraint | - | - | - | - |
| NSGA II | 1.12 | 0.47 | 1.46 | 1.49 |

Table 16: Output Results of NSGA II Algorithm for the problem No.5

| Method/metric | MID | SM | DM | SAW |
|---|---|---|---|---|
| ε-constraint | - | - | - | - |
| NSGA II | 1.02 | 1.18 | 0.89 | 0.91 |

Also, comparison of MID, SM, and DM metrics, as well as a general comparison of the two algorithms performances are demonstrated in the graphs of figures 11 to 14. According to tables (12) to (16) and graphs in figures 10 to 13 it is evident that in small scale (problem 1 to 3), to the extent that exact solving the problems in 3200 seconds timeframe is possible, the results for NSGA II algorithm and the ε-constraint Method are very close. Accordingly, the meta-heuristic algorithm used in this study results in solutions with close proximity to the exact solutions, and thereby it can be considered as a proper tool for problem solving in the cases that exact solutions are ineffective. For example, starting from Problem No. 4, the ε-constraint Method is not capable to solve problems within the timeframe exactly. As such, the NSGA II algorithm is used for solving problems on the same aforementioned scale, on account of the adequate performance of this algorithm.

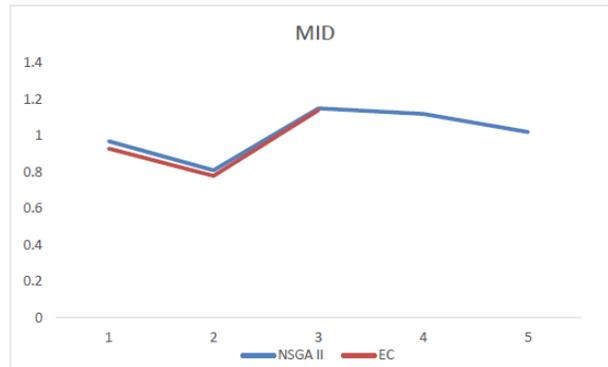

Figure 10: The values of mean distance from the ideal solution metric for the two algorithms.

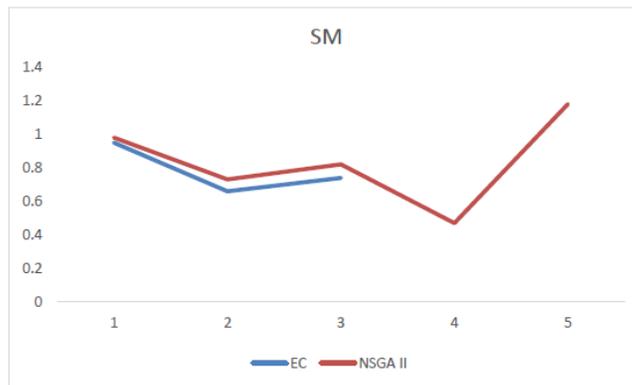

Figure 11: The values of spacing metric for the two algorithms.

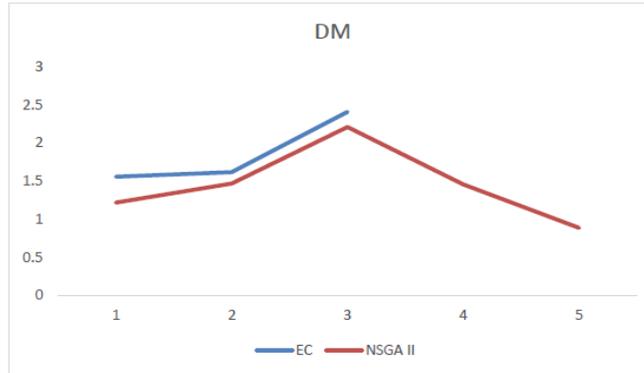

Figure 12: The values of diversity metric for the two algorithms.

The graphs of the spacing metric and the mean distance from the ideal solution, distance and diversity metrics, leads us to same conclusion that the NSGA II algorithm has an adequate performance for all of the metrics and close proximity to the exact method, which accordingly its final performance will be as close. The final comparison of the two methods is shown in figure 14 which confirms this conclusion.

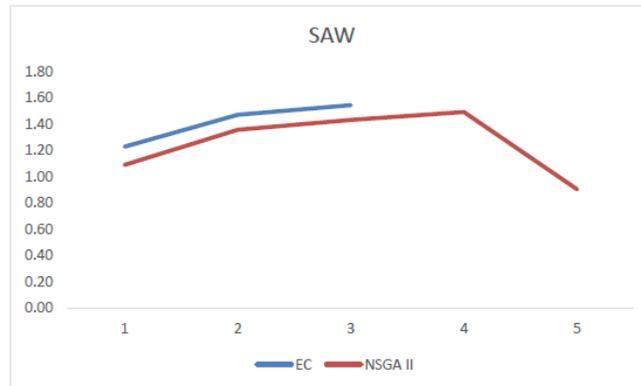

Figure 13: The final comparison of the two algorithms

According to Figure 14, this important conclusion can be confirmed that the NSGA II algorithm output is reliable and that the resulted solutions are adequately valid. Thereby this algorithm can be used as an adequate problem solving method for high scale problems. Also, the running time for solving different levels of the problems is shown in table (17).

Table 17: The running time for problem solving (seconds)

| Method/level | No.1 instance level | No.2 instance level | No.3 instance level | No.4 instance level | No.5 instance level |
|---|---|---|---|---|---|
| ε-constraint | 2.6 | 49.2 | 412.8 | 3600 | 3600 |
| NSGA II | 112.4 | 241.2 | 463.7 | 754.9 | 921.5 |

According to table (17), as the scale of the problem grows the required time for exact solution of the problem increased dramatically to the extent which after problem instance No.3, the ε-constraint Method is unable to keep up with the timeframe set finds exact solution of the problem. Instead, the meta-heuristic algorithm is able to solve the problem in much shorter time. Therefore, the NSGA II algorithm has a justifiable performance during the proper timeframe set for solving the problems. It can also be observed in figure 14.

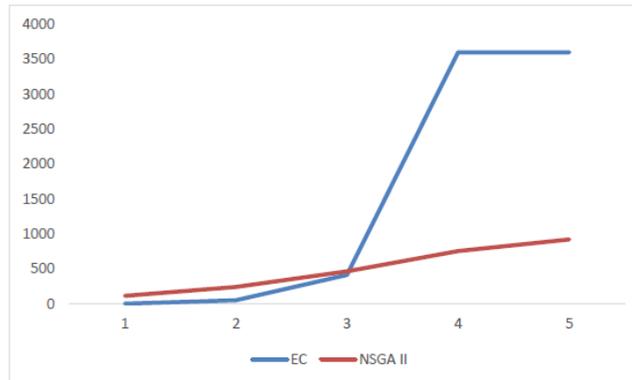

Figure 14: Timeframe for both EC and NSGA II solving methods (seconds).

Ultimately, given the adequate performance of the NSGA II algorithm, large-scale problems in this research are solved using this algorithm and, as an example of large-scale problems, the optimal Pareto-frontiers resulted using the NSGA II algorithm for problem No.5 are presented in figures 15 to 17.

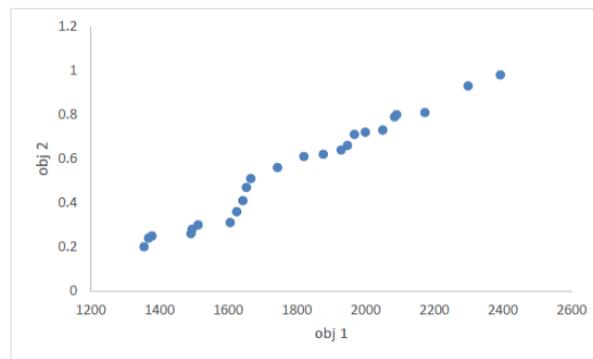

Figure 15: The Pareto-frontiers resulted for the first and the second objective functions using the NSGA II

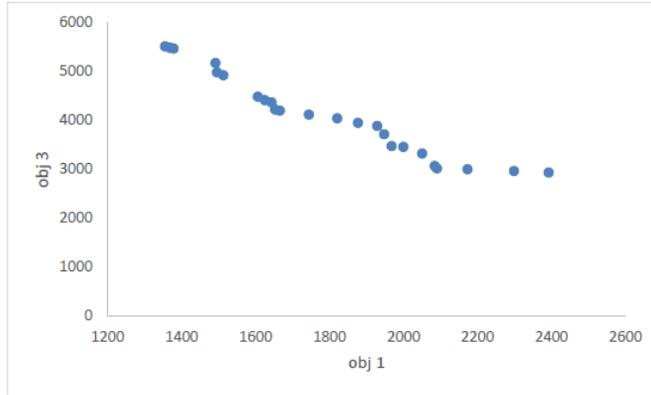

Figure 16: The Pareto-frontiers resulted for the first and the third objective functions using the NSGA II

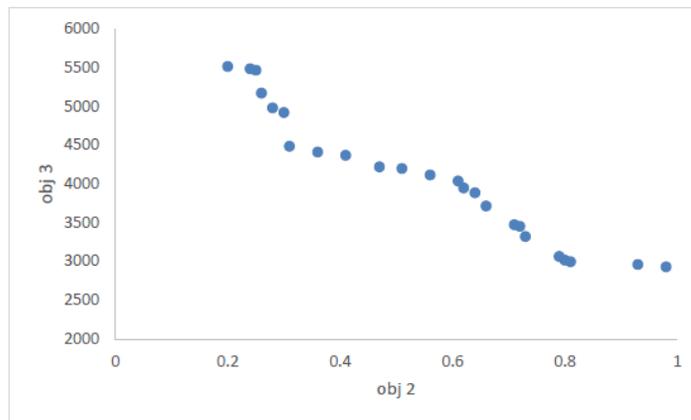

Figure 17: The Pareto-frontiers resulted for the second and the third objective functions using the NSGA II

Henceforth, the definition of the NSGAII algorithm with non-dominant sorting, an evolutionary multi-objective optimization algorithm, was discussed. In NSGAII algorithms and NSGAII based algorithms, each chromosome represents a point in the solution space and it is a feasible solution to the problem. To demonstrate a feasible solution in the proposed algorithm, a string of one and zero with the length of I*K is used, which I represent the same number of activities and K is the number of modes for each corresponding activity. And in case of selecting the mode k for activity i, the bit associated to that chromosome is zero and, otherwise it is one. Also the fitness function for the objective functions was defined. Considering 3 objective functions simultaneously for the problem, after cost calculation in the string was performed for once, then time and cost and distance constraints were imported into the three objective NSGAII algorithm, the Pareto solution are shown in the tables, which objectives contradiction is evident. In another method, three objectives counterbalance were solved using NSGAII algorithm and Pareto solutions associated to each solution are presented. And for each mode, the lowest cost and time is calculated.

## 5. Conclusions and Future research

In this study, first the necessity of paying attention to locating procedures in emergency situations is pointed out and an outline for the studied case of disaster relief supply chain was discussed and the problem was validated using GAMS and Matlab software at small scale. On the other hand, to solve this kind of problems involving three objective functions and complicated time calculation, meta-heuristic methods which yield almost optimum solutions in less time are applied. The EC method and NSGA II algorithm are among the evolutionary multi-objective optimization algorithms applied in this case. In this study the aforementioned algorithm is used for solving problems at large scale. One of the drawbacks with this model, is the required complicated calculations, and among its advantages is that it involves three objectives, including the cost of relief supplies shortages, costs of the supply's transportation, and the costs of constructing new distribution centers and finally the response time to access to people affected by the disaster which is innovative in compare to other models. The directions that can be considered for the future research are as follow:

1. Considering the subject of routing and scheduling of transportation vehicles
2. The reliability of solution
3. Extending the scope of international relief logistics
4. Proposing comprehensive models that considers the integrated multiple stages of a disaster
5. Proposing models that include destruction and disruption in relief routes.

**References:**

Afshar, M, (2011). A Mathematical Framework For Optimizing Disaster Relief Logistics. in Department of Civil and Environmental Engineering.vol. Thesis for Degree Doctor of Philosophy: University of Maryland.

Afshar, A., & Haghani, A. (2012). Modeling integrated supply chain logistics in real-time large-scale disaster relief operations. Socio-Economic Planning Sciences, 46(4), 327-338. doi:https://doi.org/10.1016/j.seps.2011.12.003

Altay, N., & Green, W. G. (2006). OR/MS research in disaster operations management. European Journal of Operational Research, 175(1), 475-493. doi:https://doi.org/10.1016/j.ejor.2005.05.016

Balcik, B., & Beamon, B. M. (2008). Facility location in humanitarian relief. International Journal of Logistics Research and Applications, 11(2), 101-121. doi: https://doi.org/10.1080/13675560701561789

Bashiri, M., & Hosseininezhad, S. J. (2009). A fuzzy group decision support system for multifacility location problems. The International Journal of Advanced Manufacturing Technology, 42(5), 533-543. doi: https://doi.org/10.1007/s00170-008-1621-3

Beraldi, P., & Bruni, M. E. (2009). A probabilistic model applied to emergency service vehicle location. European Journal of Operational Research, 196(1), 323-331. doi:https://doi.org/10.1016/j.ejor.2008.02.027

Boloori Arabani, A., & Farahani, R. Z. (2012). Facility location dynamics: An overview of classifications and applications. Computers & Industrial Engineering, 62(1), 408-420. doi:https://doi.org/10.1016/j.cie.2011.09.018

Bozorgi-Amiri, A., Jabalameli, M. S., & Mirzapour Al-e-Hashem, S. M. J. (2013). A multi-objective robust stochastic programming model for disaster relief logistics under uncertainty. OR Spectrum, 35(4), 905-933. doi: https://doi.org/10.1007/s00291-011-0268-x

Brandeau, M. L., & Chiu, S. S. (1989). An Overview of Representative Problems in Location Research. Management Science, 35(6), 645-674. doi: https://doi.org/10.1287/mnsc.35.6.645

Brimberg, J., & Juel, H. (1998). A bicriteria model for locating a semi-desirable facility in the plane. European Journal of Operational Research, 106(1), 144-151. doi:https://doi.org/10.1016/S0377-2217(97)00251-8

Camacho-Vallejo, J.-F., González-Rodríguez, E., Almaguer, F. J., & González-Ramírez, R. G. (2015). A bi-level optimization model for aid distribution after the occurrence of a disaster. Journal of Cleaner Production, 105, 134-145. doi:https://doi.org/10.1016/j.jclepro.2014.09.069

Farahani, R. Z., Hassani, A., Mousavi, S. M., & Baygi, M. B. (2014). A hybrid artificial bee colony for disruption in a hierarchical maximal covering location problem. Computers & Industrial Engineering, 75, 129-141. doi: https://doi.org/10.1016/j.cie.2014.06.012

Farahani R.Z., Hekmatfar M, (2009). Facility Locations: Concepts, Models, Algorithms and Case Studies, Physica-Verlag. doi:https://doi.org/10.1007/978-3-7908-2151-2

Jia, H., Ordóñez, F., & Dessouky, M. M. (2007). Solution approaches for facility location of medical supplies for large-scale emergencies. Computers & Industrial Engineering, 52(2), 257-276. doi:https://doi.org/10.1016/j.cie.2006.12.007

Khumawala, B. (1972). An Efficient Branch and Bound Algorithm for the Warehouse Location Problem. Management Science, 18(12), B718-B731. Retrieved July 21, 2021, from http://www.jstor.org/stable/2629558

Levin, Y., & Ben-Israel, A. (2004). A heuristic method for large-scale multi-facility location problems. *Computers & Operations Research, 31*(2), 257-272. doi:https://doi.org/10.1016/S0305-0548(02)00191-0

Louveaux, F. V. (1993). Stochastic Location Analysis: Location Science. *Location Science, 1*, 127-154.

Louveaux, F. V., & Peeters, D. (1992). A Dual-Based Procedure for Stochastic Facility Location. Operations Research, 40(3), 564-573. doi: https://doi.org/10.1287/opre.40.3.564

Mete, H. O., & Zabinsky, Z. B. (2010). Stochastic optimization of medical supply location and distribution in disaster management. International Journal of Production Economics, 126(1), 76-84. doi:https://doi.org/10.1016/j.ijpe.2009.10.004

Murali, P., Ordóñez, F., & Dessouky, M. M. (2012). Facility location under demand uncertainty: Response to a large-scale bio-terror attack. Socio-Economic Planning Sciences, 46(1), 78-87. doi:https://doi.org/10.1016/j.seps.2011.09.001

Nozari, H., Najafi, E., Fallah, M., & Hosseinzadeh Lotfi, F. (2019). Quantitative Analysis of Key Performance Indicators of Green Supply Chain in FMCG Industries Using Non-Linear Fuzzy Method. *Mathematics, 7*(11), 1020. Retrieved from https://www.mdpi.com/2227-7390/7/11/1020

Nozari, Hamed, and Agnieszka Szmelter, eds. Global Supply Chains in the Pharmaceutical Industry. Hershey, PA: IGI Global, 2019. doi: https://doi.org/10.4018/978-1-5225-5921-4

Saameño Rodríguez, J. J., Guerrero García, C., Muñoz Pérez, J., & Mérida Casermeiro, E. (2006). A general model for the undesirable single facility location problem. *Operations Research Letters, 34*(4), 427-436. doi:https://doi.org/10.1016/j.orl.2005.07.007

Tzeng, G.-H., Cheng, H.-J., & Huang, T. D. (2007). Multi-objective optimal planning for designing relief delivery systems. *Transportation Research Part E: Logistics and Transportation Review, 43*(6), 673-686. doi:https://doi.org/10.1016/j.tre.2006.10.012

Webber, A. (1999). *Theory of the Location of Industries* Uber den Standort der Industrien, Tübingen University of Chicago Press.

Yi, W., & Kumar, A. (2007). Ant colony optimization for disaster relief operations. *Transportation Research Part E: Logistics and Transportation Review, 43*(6), 660-672. doi:https://doi.org/10.1016/j.tre.2006.05.004